# A double-deletion method to quantifying incremental binding energies in proteins from experiment. Example of a destabilizing hydrogen bonding pair.

[1,2]Luis A. Campos, [1,3]Santiago Cuesta-López, [1,2]Jon López-Llano, [1,3]Fernando Falo and [1,2]Javier Sancho[*]

[1]*Biocomputation and Complex Systems Physics Institute*
[2]*Dept. Bioquímica y Biología Molecular y Celular*
[3]*Dept. Física de la Materia Condensada.*
*Fac. Ciencias. Univ. Zaragoza. 50009-Zaragoza (Spain)*

**Running Title**: Incremental binding energy of an H-bond

[*] **Correspondence to**: Javier Sancho. E-mail: jsancho@unizar.es.

**Summary:** The contribution of a specific hydrogen bond in apoflavodoxin to protein stability is investigated by combining theory, experiment and simulation. Although hydrogen bonds are major determinants of protein structure and function, their contribution to protein stability is still unclear and widely debated. The best method so far devised to estimate the contribution of side-chain interactions to protein stability is double-mutant-cycle analysis, but the interaction energies so derived are not identical to incremental binding energies (the energies quantifying net contributions of two interacting groups to protein stability). Here we introduce double-deletion analysis of 'isolated' residue pairs as a means to precisely quantify incremental binding. The method is exemplified by studying a surface-exposed hydrogen bond in a model protein (Asp96/Asn128 in apoflavodoxin). Combined substitution of these residues by alanines slightly destabilizes the protein, due to a decrease in hydrophobic surface burial. Subtraction of this effect, however, clearly indicates that the hydrogen-bonded groups in fact destabilize the native conformation. In addition, Molecular Dynamics simulations and classic double-mutant-cycle analysis explain quantitatively that, due to frustration, the hydrogen bond must form in the native structure because, when the two groups get approximated upon folding their binding becomes favorable. We would like to remark two facts: that this is the first time the contribution of a specific hydrogen bond to protein stability has been measured from experiment, and that more hydrogen bonds need to be analyzed in order to draw general conclusions on protein hydrogen bonds energetics. To that end, the double deletion method should be of help.

# Introduction

The energetics of biological macromolecules is a central unsolved problem of modern Biology that is at the core of highly important phenomena such as protein folding, and protein/ligand recognition. Understanding the contribution to protein stability of the various weak interactions between residues that appear in native protein structures will bring insight into the protein folding problem and will help to rationally tailor protein stability (Matsumura et al., 1989; Fersht and Serrano, 1993; Honig and Yang, 1995; Pace et al., 1996; Richards, 1997; Perl et al., 2000; Sanchez-Ruiz and Makhatadze, 2001). Despite advances in recent years, many fundamental questions remain unanswered. As a chief example, it is still unclear whether the ubiquitous hydrogen bonds contribute to protein stability. Conflicting views are easily held on the matter because no available technique can measure the net contribution of any two interacting groups to protein stability (the so-called incremental binding energy (Fersht et al., 1992)). Usually, estimations of the contribution of a given interaction to protein stability are based either in side chain deletion experiments or in double mutant cycle analysis. It is clear, however, that simple side chain deletion experiments aimed at breaking a given interaction and compare wild type and mutant stabilities are not informative because, in most cases, additional interactions are disrupted within the protein (Fersht, 1987; Yang and Honig, 1995). The double-mutant method (Carter et al., 1984; Horovitz et al., 1990) was conceived to alleviate this problem and, although it doesnt measure incremental binding energies, it allows to determine an interaction energy between the two side chains that, for hydrogen bonds, represents a maximum value for the contribution of the interacting groups to protein stability (Fernandez-Recio et al., 1999). In this way, double mutant cycle analysis provides upper limit values for the incremental binding energy. The problem is that, since the various interaction energies so far measured are typically small (0.0 to -1.0 kcal mol$^{-1}$), the actual stabilizing or destabilizing contribution of the bonded groups depends heavily on the value of the solvation energies of polar atoms, which can not be determined easily.

To solve this problem, we introduce here a different approach, that we term double-deletion analysis. This method focuses on pairs of interacting residues that, beyond their β-carbons, do not establish contacts with other protein residues. We show that, when two such residues are simultaneously replaced by alanines, the stability difference of the wild-type and double mutant protein, properly corrected for small differences in buried hydrophobic area, equals the so-called incremental binding energy. We then apply this double deletion analysis to quantify, for the first time, the incremental binding energy associated to a pair of surface exposed hydrogen bonded groups in a model protein. Our results, that certainly are not claimed to represent all types of hydrogen bonds in proteins, clearly show that some protein hydrogen bonds destabilize the native conformation. Using classic double mutant cycle analysis and Molecular Dynamics simulations, we discuss why they are formed, nevertheless.

## Results and discussion

**Identification of suitable hydrogen-bonded "isolated pairs"**

Hydrogen bonding is the most important single interaction in proteins, governing protein architecture and function (Jeffrey and Saenger, 1991; Branden and Tooze, 1998; Desiraju and Steiner, 1999; Lesk, 2000). For this reason we have chosen to implement the first application of the double-deletion analysis outlined in the Theory to quantify the incremental binding energy of a hydrogen bond. To find suitable candidates, we have examined several small model proteins used for stability studies: flavodoxin (1ftg), ferredoxin (1fxa), lysozyme (193l), barstar (1a19), ferredoxin-NADP+ reductase (1que), pepsin (4pep), CheY (1ehc), and cytochrome c (1crc). For each protein, we have identified all the side chain/side chain hydrogen bonds (18, 7, 12, 7, 28, 28, 2 & 2, respectively) and selected the side chains that only form one hydrogen bond (4, 3, 1, 5, 8, 5, 0 & 0, respectively). Then, we have calculated their overall solvent exposures. For the pairs with at least one residue with solvent exposure higher than 50 % (1, 1, 0, 2, 1, 1, 0, & 0, respectively) atom exposures were calculated with MOLMOL (Koradi et al., 1996), and pairs where the exposure of the atoms that would be removed upon mutation to Ala was greater than 50 % were selected. Only two pairs remained: one in ferredoxin NADP$^+$ oxidorreductase and one in flavodoxin. Of these, only the flavodoxin pair satisfied the "isolation" requisite of double deletion analysis: that the residues involved in the interaction analyzed do not make contact with any other residue in the protein beyond their Cβ. The pair is formed between the D96 and N128 side chains of the apoflavodoxin from *Anabaena* PCC 7119 (Genzor et al., 1996a) (figure 3a). The pair is also present in the holo form of the protein (Rao et al., 1992) and in three mutant flavodoxins, one in the apo form and two in the holo form, previously reported in our laboratory (Lostao et al., 2000; Lostao et al., 2003). The "isolation" requisite, itself, makes unlikely that the combined substitution of the two residues by alanines cause any significant structural rearrangement in the protein because no additional residue loses or gains interactions upon mutation.

As outlined above, the identification of suitable candidates has been carried out by visual inspection. The fact that only one in 104 side chain/side chain hydrogen bonds analyzed has turned out to be appropriate suggests that an automated analysis of the Protein Data Bank would be helpful to find candidates to analyze other side chain interactions and other types of hydrogen bonds. Work is in progress in this direction (F. Pazos, personal communication).

**Integrity of the mutant proteins**

The overall integrity of the D96A and N128A single mutants and of the D96A/N128A double mutant has been initially assessed by comparing the fluorescence, far-UV CD, and near-UV CD spectra to those of wild type. The fluorescence emission (not shown) and the far and near-UV CD spectra (figure 4) of the three mutant proteins are almost identical to those of the wild type protein. In addition to maintaining the overall fold, double deletion analysis requires, as double mutant cycle analysis does, that the local protein structure is not altered by the mutations introduced. Although the x-ray structures of the mutants are not available (they have failed to crystallized) there is firm crystallographic evidence, coming from the structure of a highly related flavodoxin, that the implemented mutations to alanine do no cause local perturbations. As shown in figure 3b superimposed to the structure of the wild type *Anabaena* apoflavodoxin (Genzor et al., 1996a), the flavodoxin from *Chondrus Crispus* (2fcr) (Fukuyama et al., 1992) contains an aspartic residue (D100) that is structurally equivalent to the D96 in *Anabaena* apoflavodoxin. However, at the position equivalent to N128, the *C. Crispus* flavodoxin displays a glutamate (E132), and therefore hydrogen bonding with its D100 neighbor is not possible. In this respect, and given that E132 and D100 should repel each other due to their charges, the *C. Crispus* flavodoxin exemplifies the structural consequences of a mutation that, potentially, is much more disruptive than the D96A and N128A mutations implemented here. Yet, as figure 3c shows, the *C. Crispus* flavodoxin accommodates the mutation by simply rotating the glutamate side chain so that the carboxyl group points to the solvent. The $C_\beta$ of the *C. Crispus* E132 is at the same position as that of *Anabaena* N128, and remarkably D100 remains unmoved from the position of the structurally equivalent *Anabaena* D96. If the mutation of one of the residues involved in the pair leaves the other unchanged and unpaired (beyond the $C_\beta$) it is difficult to envision that the mutations to alanine may cause any local alteration. Based in this fact, we have modeled the structure of the double mutant D96A/N128A by simply mutating *in silico* the wild type residues to alanine.

A more direct indication that the double mutation is well tolerated by the protein without significant rearrangements has been obtained from Langevin Molecular Dynamics simulations of wild type and double mutant apoflavodoxin. The simulations have been run for 4.5 ns, which has proved long enough to reach an equilibrium configuration. Figure 5a shows, for the wild type and the D96A/N128A double mutant, the RMS deviations as a function of simulation time (relative to the starting, previously minimized structures) of the residues located within a 6 Å radius of any of the atoms of the D96 carboxyl and N128 carboxamide groups. It is clear that the RMSs values hardly change with time and that no noticeable differences can be observed between the wild type and the double mutant RMS traces. In fact, when the side chains are included together with the backbone atoms in the RMSs calculations, the wild type trace is somewhat less stable than that of the double mutant, as a consequence of the dynamics of the hydrogen bonded D96 and N128 side chains. The RMS traces of the backbone atoms are very stable for the two proteins, which can hardly be distinguished. This local stability of the loops bearing the hydrogen bonded residues correlates with their low B-factor in the x-ray structure (Genzor et al., 1996a). In fact, our simulations reveal larger departures from the x-ray structure in distant regions of the protein that display high B-factors in the crystal (not shown).

**Incremental binding energy of a hydrogen bond**

To calculate the incremental binding energy of the mutated bond, the stability of wild type and D96A/N128A apoflavodoxins has been measured by urea denaturation (figure 6a) as described (Genzor et al., 1996b). To avoid analysis complications arising from long range electrostatic interactions, we have performed all measurements in the presence of 0.5 M NaCl. Previous work (Maldonado et al., 2002) has shown that this salt concentration effectively masks medium and long range electrostatic interactions in apoflavodoxin. As indicated by its urea concentrations of mid denaturation (Table 1), the double mutant lacking the hydrogen bond is slightly less stable than the wild type protein by $0.21 \pm 0.06$ kcal mol$^{-1}$ (or $0.29 \pm 0.23$ kcal mol$^{-1}$, if less accurate individual *m* values are used instead of an average *m* value, see methods). This difference in stability (wild type minus double mutant), that we have termed double-deletion energy, equals the contribution of the two hydrogen bond forming groups to protein stability (relative to having two alanines) plus a solvation term (see equations 3, 4 and 5). As indicated in the Theory, the solvation term in equation 5 concerns essentially apolar atoms, its sign is known and its actual value can be calculated from empirical equations with reasonable accuracy. From the differential solvation in the folded state (-18.7 Å$^2$ corresponding to the apolar atoms neighboring the 96 and 128 side chains in the wild type and in the modeled double mutant proteins (Table 2) plus -62.5 Å$^2$, corresponding to the D96 and N128 Cβ atoms (Table 3)), together with the differential solvation in the unfolded state (-58.4 Å$^2$, corresponding to the D96 and N128 Cβ atoms (Table 3)), the overall apolar area change related to the solvation term in equation 5 amounts to -22.8

Å$^2$. We thus calculate (equation 7) this solvation term at -0.65 kcal mol$^{-1}$. On the other hand, a more accurate calculation of this term can be performed if, instead of considering the solvent exposures in the wild type crystal structure and in the double mutant model, averages of the exposure to solvent of the two proteins during the 4.5 ns Molecular Dynamics trajectories are used (see Table 2). From the averaged exposures of each protein, we calculate that the overall differential area (wild type minus double mutant) exposed to solvent is of -14.5 Å$^2$, rather that -22.8 Å$^2$, which sets the solvation term in equation 5 at -0.42 kcal mol$^{-1}$. An estimation of the error associated to the solvation term can be obtained if an average (±SE) of the areas calculated by the two methods is considered (18.7 ± 4.2 Å$^2$) and the uncertainty in the multiplying constant on equation 7 is taken into account. For the constant (28.7 cal mol$^{-1}$Å$^{-2}$, see methods) we calculate a standard error of ± 2.8 cal mol$^{-1}$Å$^{-2}$ from the values of eleven different factors proposed in recent years (see methods). In this way, the solvation term is estimated at -0.54 ± 0.13 kcal mol$^{-1}$.

The net contribution of the hydrogen bonding carboxylate and carboxamide groups of D96 & N128 to protein stability can now be calculated (see equation 5) by combining the experimentally determined double deletion energy and the solvation term, and it turns out to be of +0.33±0.14 kcalmol$^{-1}$ (using non-averaged *m* slopes a less accurate quantification can be offered at +0.25±0.26 kcalmol$^{-1}$). The contribution of the two hydrogen bonding groups is thus, in principle, small and destabilizing! However, we would like to point out that the destabilizing contribution of the D96/N128 bonding groups to protein stability could be larger. This is so because we have used in our calculations solvent exposed areas in the denatured state that are based in the exposures observed in model tripeptides, and therefore could be unrealistically large. As more accurate determinations of solvent exposures in denatured states are being performed (by averaging states populated in Molecular Dynamics simulations and by considering longer model peptides) the reported values of solvent exposures tend to shrink. We note, in this respect, that if the actual exposures of the C$\beta$ in the denatured state were smaller than those used in our calculations, and reported in Table 3, by the amount suggested by Creamer and coworkers (Creamer et al., 1995) for longer peptides, the solvation term in equation 5 would still be negative but significantly larger that the estimated -0.54 ± 0.13 kcal mol$^{-1}$ (actually it would amount to around -1.3 kcal mol$^{-1}$). In this more realistic scenario, the destabilizing contribution of the D96/N128 hydrogen bond would be of around +1.1 kcal mol$^{-1}$.

The reason why the wild type protein is slightly more stable that the double alanine mutant is that a significant stabilization is obtained from an increased hydrophobic effect arising from the shade cast by the carboxylate and carboxamide groups of D96 and N128 on neighboring apolar groups, not directly in contact. This effect does not stabilize the hydrogen bond itself because it would arise to a similar extent in the wild type protein if the hydrogen bond were not formed.

In agreement with our finding of a destabilizing contribution of hydrogen bonding groups to protein stability, there is recent work by several laboratories that also points to a destabilizing contribution of hydrogen bonding groups in proteins (Ma and Nussinov, 2000; Guerois et al., 2002). The same view is held by detailed calculation (Ben-Tal et al., 1997) and measurement (reviewed in Ben-Tal et al., 1997) of the dimerization energy of model compounds. The contrasting view supporting a stabilizing contribution of hydrogen bonding groups to protein stability based in the analysis of single deletion experiments has been reviewed by Myers and Pace (Myers and Pace, 1996). In our view, single-deletion experiments are unlikely to clarify so subtle a matter, among other things because, as it is acknowledged by Myers and Pace, "we are left to guess at the hydrogen bonding status of the remaining partner".

**Why a destabilizing interaction is established**

It may seem paradoxical that a destabilizing interaction like this hydrogen bond is present at all in the native structure. The paradox, however, can be easily explained in a quantitative manner by conceptually dividing the folding of the protein into two processes (figure 2). First, the protein folds to a virtual intermediate where residues $i$ and $j$ are close in space but do not yet interact with each other. In the second step, the $i$ and $j$ side chains approach and form a bond. It is the free energy difference of the second step ($\Delta G_{II}$) what governs the stability of the hydrogen bond in the context of the native structure and the fact that the hydrogen bond is observed in the crystal structure merely suggests that $\Delta G_{II}$ should be negative. To test this interpretation we have quantitated $\Delta G_{II}$ both from experiment and from simulation.

The experimental approach relies in the similarity of the solvation energies of $i$ and $j$ residues in the virtual intermediate depicted in figure 2 and those displayed by the same residues in the single mutants (i0 and 0j) present in the double mutant cycle scheme (figure 1). Although identifying one set of solvation energies with the other is a simplification (because it is likely that the i and j residues would be more desolvated in the virtual intermediate than in the single mutants) it provides a useful way to estimate $\Delta G_{II}$ from classical double mutant cycle analysis. Assuming that the solvation energies of the i and j residues in the single mutants approximate those in the virtual intermediate, the interaction energy measured by double mutant cycle represents the binding energy of the $i$ and $j$ residues interacting from the close to native intermediate state ($\Delta G_{II}$ in figure 2) plus a solvation term: $\Delta\Delta G_{Pw}$ (ij – i0 - j0 – 00) that essentially refers to apolar surface and can be estimated independently. We have thus resorted to double mutant analysis, prepared the two related single apoflavodoxin mutants, and determined their stability by urea denaturation (figure 6b). The double mutant cycle-derived interaction energy is of –0.19 ± 0.06 kcal mol$^{-1}$ (Table 1; or, less accurately, -1.3±0.7 kcal mol$^{-1}$, if individual instead of averaged *m* values are used). Since the solvation term amounts in this case to +0.32 ± 0.03 kcal mol$^{-1}$ (11.0 Å$^2$, Table 2), $\Delta G_{II}$ is calculated at –0.51 ± 0.07 kcalmol$^{-1}$: stabilizing! (a larger, but less accurate value of -1.6 ± 0.7 kcal mol$^{-1}$ would be calculated from individual *m* slopes).

In fact, due to the expected greater desolvation of the side chains in the virtual intermediate than in the single mutants, and due to the smaller entropy change of bond formation in the intermediate than in the unfolded state, the calculated value of $\Delta G_{II}$ = -0.51 kcal mol$^{-1}$ underestimates the binding energy of the hydrogen bond within the folded structure. We believe a more accurate determination of $\Delta G_{II}$ can be achieved by careful analysis of Molecular Dynamics simulation of the wild type protein. To that end, we have specifically monitored the dynamics of the D96/N128 bond. The bond can be established by either of the OD1 and OD2 oxygen atoms of the D96 side chain, and, indeed, the alternative involvement in the bond of the two oxygens is observed (not shown). To describe the energetics of a carboxylate/carboxyamide hydrogen bond, the two configurations of the bond should not be differentiated. Monitoring the distances between the D96/N128 residues during the 4.5 ns trajectory reveals that, in addition to the swapping of oxygens, the bond breaks and reforms many times during the sampled trajectory. In some cases, the Asp side chain is observed to bend into the solvent where it establishes new bridges with bulk water molecules. To illustrate the dynamics of the bond, the shortest of the distances between the N128 side chain H atom and any of the D96 OD1 and OD2 atoms is show in figure 5b as a function of time. Some clear breaking events are evident in the trajectory. The fluctuation of O-H distances around the equilibrium position is best observed in the histogram shown in figure 5c, where two regions can be distinguished: a narrow peak centered around the equilibrium bond distance (1.8 Å) and a very broad distribution from 2.5 Å to around 8 Å corresponding to the unbound configuration. This is consistent with local two-state behavior and allows quantification of the binding energy of bond formation from the folded state. Using a typical 2.5 Å threshold as the bond breaking O-H distance, we calculate that the hydrogen bond remains formed 85% of the time, which reflects a binding energy of -1.0 ± 0.1 kcal mol$^{-1}$ (allowing for a 0.1 Å error in the threshold). As was anticipated above, this value of $\Delta G_{II}$ is larger than the one calculated from the double mutant cycle approximation (-0.51 ± 0.07 kcal mol$^{-1}$) and we consider it to be more accurate. Whatever the exact value of $\Delta G_{II}$, both the experimental analysis and the Molecular Dynamics simulation clearly indicate that forming the hydrogen bond from the compact, partly desolvated, close to native state does indeed significantly stabilize the protein. The paradox is thus solved as follows: adding to the apoflavodoxin polypeptide two hydrogen bonding groups (the carboxyl and carboxamide in D96 and N128) that form a hydrogen bond in the native state destabilizes the native protein, and yet the two groups are forced to interact and form the bond because, in the context of the folded protein, bond formation becomes favorable. Why this is so in this particular case is open for interpretation and it is clear from the Molecular Dynamics simulations that the hydrogen bond can be broken by side chain rotations. We point out that two potential contributions to the stability of the hydrogen bond in the context of the native structure could be a lower effective concentration of water felt by the interacting residues in the folded state (as compared to the unfolded state) and a reduced entropy change of binding in the native state due to their proximity and to the fact that the side chain of N128 is relatively constrained. Whatever the specific cause, which is difficult to precise, it seems that frustration manifesting in protein folding may similarly drive the formation of other non-stabilizing or even destabilizing interactions that will thus be present in native proteins. Recent work on a salt bridge also points to this direction (Luisi et al., 2003). Thus, statistical potentials derived from contact frequencies in proteins do not necessarily reflect the energetics of pair-wise interactions, if the denatured state is taken as the reference.

**Concluding remark**

It is tempting to extrapolate the finding that the hydrogen bond analyzed here displays a positive (destabilizing) incremental binding energy to conclude that hydrogen bonds destabilize proteins or at least do not stabilize them. Indeed, our finding agrees well with the fact that, as far as we know, no claims of protein stabilization have been made based in engineering pairs of polar groups to form new hydrogen bonds, which suggests that perhaps proteins cannot be stabilized in this way. It is clear, however, that more hydrogen bonds must be studied to establish whether the picture offered by this Asp/Asn bond can be generalized. This is so because the differential solvation energies ($G_{iw}$, $G_{jw}$) will vary with solvent exposure in the native state, and because their values for the various polar groups appearing in proteins are different (Jeffrey and Saenger, 1991), as are different the intrinsic strengths of the bonds they establish ($G_{ij}$). Therefore, a surface Asp/Asn hydrogen bond may be significantly different from a surface bond involving other residue types or from a buried Asp/Asn bond. In terms of overall protein energetics, it would be particularly interesting to see what the trend is for carefully chosen buried hydrogen bonding groups, as they could report on the contribution of the ubiquitous main chain hydrogen bonds to protein stability. A recent study suggests, from isotope effect measurements, a different contribution to protein stability for main chain hydrogen bonds located in α-helices and in β-sheets (Shi et al., 2002), which stresses the subtlety of the balance.

The more important conclusion of this work is that the double deletion method offers an experimental way to quantify precisely the contribution of side chain interactions to protein stability. However, it requires a very demanding selection of suitable interacting pairs that makes unlikely to find, in a particular model protein, more that one useful pair to investigate a given interaction. The method therefore has both advantages and disadvantages compared to double mutant cycle analysis.

## Materials and Methods

**Theory of double deletion analysis**. Let *i* and *j* be interacting residues in a protein. In well-chosen cases, where the individual and the simultaneous replacement of these groups by alanines does not alter the local protein structure, a double mutant cycle can be constructed with the wild type, single and double mutants so that an *ij* interaction energy is calculated from the conformational stabilities of the four proteins (Fersht et al., 1992).

$$\Delta G_{int} = \Delta G_{wt} - \Delta G_{i0} - (\Delta G_{0j} - \Delta G_{00}) \qquad [1]$$

where $\Delta G_{wt}$, $\Delta G_{i0}$, $\Delta G_{0j}$ and $\Delta G_{00}$ are the stabilities of the wild type, the *j*→Ala, the *i*→Ala, and the double mutant protein, respectively. Using energy inventories (figure 1), it can be shown that, for non-disruptive mutations, the interaction energy is made of the following terms (all relative to the unfolded state):

$$\Delta G_{int} = G_{ij} + G_{iw}(ij) + G_{jw}(ij) - G_{iw}(i0) - G_{jw}(0j) + \Delta\Delta G_{Pw}(ij - i0 - j0 - 00) \quad [2]$$

where $G_{ij}$ refers to the specific interaction of the two residues, $G_{iw}(ij)$ and $G_{jw}(ij)$ are the solvation energies of the two residues in the wild type protein, $G_{iw}(i0)$ and $G_{jw}(0j)$ are the solvation energies of each residue in the single mutant proteins, and $\Delta\Delta G_{Pw}$ (ij – i0 – j0 – 00) summarizes the changes in the solvation of the rest of the protein in the four proteins.

Suppose now, that the $i$ and $j$ interacting residues do not contact, beyond their β carbons, any other residue in the protein. If, in addition, long range electrostatic interactions are masked by working at high ionic strength, the interaction of the $i$ and $j$ residues with the rest of the protein (relative to that of alanine) is zero. Thus, if the conformational stability of the double mutant is subtracted from that of the wild type protein, a "double deletion" energy ($\Delta G_{dd}$) is obtained that, according to the energy inventory (figure 1), equals:

$$\Delta G_{dd} = G_{ij} + G_{iw}(ij) + G_{jw}(ij) + \Delta G_{Pw}(ij - 00) \quad [3]$$

On the other hand, the contribution to protein stability (relative to two alanines) of a pair of residues that interact in the native conformation is given by the incremental binding energy ($\Delta G_b$), defined as (Horovitz et al., 1990; Fersht et al., 1992):

$$\Delta G_b = G_{ij} + G_{iw}(ij) + G_{jw}(ij) \quad [4]$$

Combining equations 3 and 4:

$$\Delta G_b = \Delta G_{dd} - \Delta G_{Pw}(ij - 00) \quad [5]$$

Equation 5 is the key to double deletion analysis because in many cases, as in the example presented in this work, the solvation term ($\Delta G_{Pw}(ij - 00)$) refers to apolar surface, and its calculation is feasible from known empirical equations (see below). It should be noted that, in $\Delta G_{Pw}(ij - 00)$, the solvation of the mutated residues in the unfolded state does not cancel out, unlike in double mutant analysis. Since both $G_{Pw}(ij)$ and $G_{Pw}(00)$ are differential solvation energies (folded minus unfolded), $\Delta G_{Pw}(ij - 00)$ can be expressed as:

$$\Delta G_{Pw} (ij - 00) = \Delta G^{fold}_{Pw}(ij - 00) - \Delta G^{unf}_{Pw}(ij - 00) \qquad [6]$$

The $\Delta G^{fold}$ term can be calculated from the surface exposed areas in the wild type and double mutant folded structures. The $\Delta G^{unf}$ term, from the exposure in the unfolded state of the beta carbons of the wild type $i$ and $j$ residues and of the alanine ones in the double mutant. As in classical double mutant cycle analysis, it is assumed that the mutated residues do not interact in the unfolded state.

**Surface calculations and quantification of solvation energies**. The double deletion method has been applied to determine the contribution to protein stability of a surface exposed hydrogen bond formed by the Asp96 and Asn128 side chains of the apoflavodoxin from *Anabaena* PCC 7117 (1ftg). To that end, the solvent accessible surface areas of the wild type and the D96A/N128A double mutant proteins have been calculated in two different ways. One uses the x-ray structure of the wild type protein and a model of the double mutant that was build by substituting the Asp and Asn residues by Ala. Solvent accessible surface area is calculated with *Naccess 2.1*.1 (Hubbard and Thornton, 1993) using a probe sphere of 1.4 Å (Lee and Richards, 1971). The other uses, as representatives of the proteins, averages of the structures obtained along Molecular Dynamics simulations (see below). Since the local RMS deviations (t minus t=0) around the hydrogen bond investigated hardly change along the simulations of the proteins, structures have been averaged that sample the entire trajectories. In this approach, average solvent accessible surface areas have been calculated using *Naccess*

*2.1.1* (Hubbard and Thornton, 1993), interfaced with CHARMM through a home made program. The surface exposed areas of the proteins, calculated by either of the two methods, have been then used to calculate the changes in solvent exposed area upon mutating D96 and N128 to Ala (excluding the mutated carboxyl and carboxamide groups, which are explicitly excluded in the $\Delta G_{Pw}$ (ij – 00) term of equation 3 because this term refers to the interactions between the rest of the protein and water).

The surface area of the beta carbons of residues D96, N128, A96 and A128 in the unfolded state have been calculated using data from molecular dynamics simulations of Ala-X-Ala tripeptides (Zielenkiewicz and Saenger, 1992). These data agree with those reported for tripeptides by Creamer and coworkers (Creamer et al., 1995; Creamer et al., 1997), who suggest, however, that in longer peptides side chain exposures are reduced to about 65 % of their values in tripeptides.

The quantification of solvation energies (in cal mol$^{-1}$) from changes in solvent area (in Å$^{-2}$) has been performed using the following relationship:

$$\Delta G_{Pw} (ij - 00)^{apolar} = 28.7(\pm 2.8) \Delta ASA(ij - 00)^{apolar} \qquad [7]$$

where the converting factor is the average (±SE) of eleven different factors proposed since 1991 (Sharp et al., 1991; Eriksson et al., 1992; Schiffer et al., 1992; Blaber et al., 1993; Pinker et al., 1993; Koehl and Delarue, 1994; Vajda et al., 1995; Eisenhaber, 1996; Weng et al., 1997). The very small change in polar area (Table 2) has not been considered. According to different parameterizations (Vajda et al., 1994; Xie and Freire, 1994), its contribution to $\Delta G_{Pw}$ (ij – 00) would be between 0.00 and 0.06 kcal mol$^{-1}$.

**Mutagenesis, protein expression and purification, and spectroscopic characterization.** PCR-mutagenesis of the *Anabaena* PCC 7119 flavodoxin gene was performed with the *QuikChange* kit and the mutations identified by sequencing.

Expression of the gene (Fillat et al., 1991) was done in *E. coli*. Purification and removal of the FMN prosthetic group was performed as described (Genzor et al., 1996b). Near-UV CD spectra (260 to 310 nm) of wild type and mutant proteins were obtained with a 1 cm cuvette and 30 µM protein solutions in 50 mM mops, pH 7. Far-UV CD spectra (200 to 250 nm) were recorded with a 1 mm cuvette, at the same protein concentration an a 5 mM mops, pH 7 buffer containing 15 mM NaCl.

**Stability measurements.** The conformational stability of the apoflavodoxin from *Anabaena* has been extensively characterized in our laboratory (Genzor et al., 1996b; Maldonado et al., 1998a; Maldonado et al., 1998b; Fernandez-Recio et al., 1999; Irun et al., 2001a; Irun et al., 2001b; Langdon et al., 2001; Maldonado et al., 2002; Lopez-Llano et al., 2004a; Lopez-Llano et al., 2004b) and its equilibrium urea denaturation has been shown to be two-state (Genzor et al., 1996b; Fernandez-Recio et al., 1999; Irun et al., 2001a; Irun et al., 2001b; Langdon et al., 2001; Maldonado et al., 2002). The stability of wild type and mutant apoflavodoxins has been measured by urea denaturation as described (Genzor et al., 1996b), but using a ratio of intensities (320/380 nm). Because *m* values are typically determined with large errors when urea unfolding curves of proteins are fitted using the linear extrapolation method (Santoro and Bolen, 1988), which is in contrast with the much greater reproducibility of denaturant concentrations of mid denaturations, protein stability differences are most accurately determined using an average *m* value for the different proteins, although this practice is sometimes questioned (Yi et al., 2003). Based in previous work in our laboratory with wild type and mutant apoflavodoxins, we have estimated (Fernandez-Recio et al., 1999) that the accuracy of stability differences between apoflavodoxin variants calculated using an average *m* slope is of around ± 0.06 kcal mol$^{-1}$ (this applies to $\Delta G_{dd}$; see equation 3), and that of stability differences between four variants at around ± 0.08 kcal mol$^{-1}$ (this applies to $\Delta G_{int}$). If, however, the individual *m* values obtained for each protein variant are used, much larger errors are obtained due to the intrinsic poor reproducibility of *m* values. In this work we report stability differences calculated using both an average *m* value and individual *m* values. The two sets of data are in qualitative agreement and point to the same conclusions. We consider the data obtained using an average *m* value as more accurate. Another potential source of inaccuracy in protein stability determinations is batch-related protein stability differences. However, in the particular case of *Anabaena* apoflavodoxin, we have not observed over the years significant differences among different batches of the wild type protein (not shown).

**Molecular Dynamics.** Molecular Dynamics simulations of the apoflavodoxin wild type structure (1ftg) and of the modelled double mutant were performed using the CHARMM (c27b2) package (Brooks et al., 1983). An initial step of minimization was applied to both structures, using several cycles of Steepest Descent, Conjugate Gradient and Adopted Basis Newton-Raphson. Solvation of the systems was achieved by placing the protein structures inside a pre-equilibrated cubic box of TIP3P water molecules (Jorgensen et al., 1983). To reduce edge effects, periodic boundary conditions were applied, and the SHAKE algorithm (Ryckaert et al., 1977) was used to hold rigid the internal geometry of the water molecules, according to the Jorgensen description (Jorgensen et al., 1983). Long-range electrostatic interactions were modelled with the particle-mesh Ewald method (Essmann et al., 1995), using a 12.0 Å cutoff and a grid spacing of about 1.0 Å. To achieve an appropriate neutralization of the system, Na$^+$ counterions were iteratively placed. Initially, they were randomly positioned, avoiding overlaps with the protein and removing the water molecules located within a 2.5 Å radius of the ions introduced. Then, a short minimization was performed, keeping the protein fixed, to improve the solvation of the ions, and a 10 ps CPT dynamics was run (298 K, 1 at) (Feller et al., 1995), to allow the solvation cage to expand in order to avoid internal strains.

Langevin Dynamics were used to heat the system and to produce trajectories in the Canonical Ensemble (Paterlini and Ferguson, 1998; Krivov et al, 2002). The use of Langevin Dynamics is cpu time-consuming (as compared to using other traditional algorithms, such as nose-Hoover) but is advantageous in that it guaranties a better representation of the ensemble. Since the aim was the determination of equilibrium properties, the choice of the friction coefficient should not affect the results (provided the fluctuation-dissipation relation is fulfilled), although it can influence the dynamics (see below). A Leapfrog Verlet integrator with a time step of 1 fs was used. The friction coefficient $\gamma$ in the Langevin equations was set to 64 $ps^{-1}$ for solvent molecules (Smith et al., 1993) and to 1.5 $ps^{-1}$ for protein atoms. This choice allows a fast equilibration of the solvent and speeds up the dynamical processes inside the protein (Zagrovic and Pande, 2003). In addition, it eliminates the spoiling high frequency modes in the solvent that do not concern our study. The simulations began with a 50 ps, slow, progressive heating to the working temperature (298 K), followed by a production run of 4.5 ns.

**Acknowledgements.** Work supported by the Spanish MCYT (BMC 2001-252, MCyT-BFM2002-00113) and DGA (P120/2001). LAC and SC-L supported by FPU fellowships and JL-L by a Bask Government fellowships. We thank P.J. Martínez for computational support.

**Table 1.** Stability of wild type and mutant proteins[a]

| Protein | $m^b$ (kcal mol$^{-1}$ M$^{-1}$) | $U_{1/2}^c$ (M) | $\Delta G^d$ (kcal mol$^{-1}$) | $\Delta G_{av}^e$ kcal mol$^{-1}$) |
|---|---|---|---|---|
| WT | 2.27 ± 0.06 | 3.270 ± 0.022 | 7.43 ± 0.16 | 7.10 ± 0.05 |
| D96A/N128A | 2.25 ± 0.05 | 3.175 ± 0.001 | 7.14 ± 0.16 | 6.89 ± 0.03 |
| D96A | 2.15 ± 0.21 | 3.181 ± 0.003 | 6.85 ± 0.66 | 6.90 ± 0.01 |
| N128A | 2.02 ± 0.06 | 3.176 ± 0.010 | 6.41 ± 0.23 | 6.90 ± 0.02 |

[a] Urea denaturation performed at 25.0 °C, in 50 mM mops, pH 7.0 with 0.5 M NaCl.

[b] Slope of a linear plot of ΔG versus urea concentration. Mean of two determinations ± standard error.

[c] Urea concentration of mid-denaturation. Mean of two determinations ± standard error.

[d] Standard free energy of unfolding calculated for each protein as $m_i$ times $U_{1/2i}$ Mean of two determinations ± standard error.

[e] Standard free energy of unfolding calculated for each protein as $m_{av}$ times $U_{1/2i}$ where $m_{av}$ = 2.17 is the average slope of all determinations. Mean of two determinations ± standard error. We consider this data to be more accurate.

Table 2. Incremented surface area per atom type in the folded state[a]

| Protein | Amino acid residues with solvent accessible surface changes relative to WT | New area exposed to solvent in the folded state by atom type (Å$^2$) | | | | |
|---|---|---|---|---|---|---|
| | | C(C) | C(O) | C(N) | O | N |
| D96A[b] | A95, A96, N97, N128, N129 | 33.5 | 0.0 | 0.0 | 0.0 | 1.4 |
| N128A[b] | A95, F127, A128, D129 | 29.9 | 3.5 | 3.3 | 2.7 | 0.0 |
| D96A/N128A[b] | A95, A96, N97, F127, A128, N129 | 74.4 | 3.5 | 3.3 | 2.6 | 1.5 |
| | | 11.9[d] | 3.5[d] | 3.3[d] | 2.6[d] | 1.5[d] |
| D96A/N128A[c] | Y94, A95, A96, N97, D126, F127, A128, N129 | 65.7 | 2.6 | 1.0 | 0.8 | 1.4 |
| | | 6.8[d] | 2.6[d] | 1.0[d] | 0.8[d] | 1.4[d] |

[a] These areas refer only to the folded state. They exclude the i and j mutated residues and are pertinent to estimate the magnitude of $\Delta\Delta G_{Pw}(ij\text{-}i0\text{-}j0\text{+}00)$ in eq. 2 because in the double mutant cycle all terms concerning solvations in the denatured state cancel out.

[b] Calculated with *Naccess 2.1.1* (Hubbard and Thornton, 1993) from the x-ray structure of the wild type protein (1FTG) and the models of the single and double mutants generated by replacing *in silico* the Asp and/or Asn side chains by Ala ones.

[c] Calculated with *Naccess 2.1.1* (Hubbard and Thornton, 1993) (from averages of the wild type and double mutant structures generated along 4.5 ns Langevin Molecular Dynamics trajectories).

[d] Excluding the β carbons of residues at 96 and 128. The data concerning carbon atoms (C(C), C(O) and C(N)) is used, together with data in Table 3, for the calculation of the solvation term in equation 6.

**Table 3**. Exposed area of $C_\beta$ atoms of residues 96 & 128 in the folded and unfolded states (Å$^2$)[a]

| Residue | Folded state | Unfolded state[b] |
|---------|--------------|-------------------|
| D96     | 18.9         | 36.3              |
| N128    | 8.1          | 38.3              |
| A96     | 48.4         | 66.5              |
| A128    | 41.1         | 66.5              |

[a] These areas allow to calculate the contribution of the beta carbons to $\Delta G_{Pw}$ (ij-00) in eq. 3. To calculate this term, the atoms that are removed by mutation are not pertinent but the solvation of their beta carbons in the folded and unfolded states has to be taken into account because, unlike in double mutant cycle calculations, it does not cancel out.

[b] Data from Zielenkiewicz and Saenger, 1992.

# Legends to figures

**Figure 1**. Energy inventory in double mutant cycle and double deletion analyses. The equations show the relationship between the incremental binding energy from the unfolded state (the contribution of any two groups to protein stability), the double mutant cycle interaction energy, and the double deletion energy. See theory.

**Figure 2**. Scheme depicting the folding of a protein as divided into two steps. In the first one, with $\Delta G_I$, the protein get folded to a virtual intermediate (essentially folded) where the i and j residues do not yet establish an interaction. Here, the solvations of the i and j residues are equivalent to those in the folded state of the 0j and i0 single mutants, and the interaction between them is considered close to zero. In the second step, with $\Delta G_{II}$, the two residues establish an interaction. The equations show the relationship between $\Delta G_{II}$, that represents the incremental binding energy from the virtual, folded, intermediate, and the interaction energy, calculated from double mutant cycle analysis.

**Figure 3.** A. Ball and stick representation of the *Anabaena* apoflavodoxin structure (1ftg) showing the hydrogen bonded residues D96 and N128. Hydrogen bonds in magenta. B. Superposition of the apoflavodoxin from *Anabaena* and holo flavodoxin from *Chondrus Crispus* (2fcr) showing the *Anabaena* hydrogen bonded residues: D96 and N128, and their structural equivalents: D100 and E132. The perfect conservation of the structure at the site of mutation in the *Chondrus Crispus* protein, where the hydrogen bond is no longer possible, can be appreciated. C. Superposition of the Anabaena apo and holo (1flv) flavodoxin structures showing the conservation of the hydrogen bond upon FMN cofactor binding.

**Figure 4**. Near-UV (a) and far-UV (b) circular dichroism spectra of wild type (solid circles), D96A (solid triangles), N128A (open circles) and D96A/N128A (open triangles) mutants. Spectra obtained at 25.0 ± 0.1 ºC in mops 50 mM pH 7.0.

**Figure 5.** Molecular dynamics simulation of wild type *Anabaena* apoflavodoxin and of the D96A/N128A double mutant. (a) RMSD of the overall structures and of the atoms within a 6 Å radius of the carboxyl and carboxamide groups removed upon mutation. The initial raising of the RMSD traces corresponds to the initial heating to 298 K of the reference minimized structures. (b) Evolution of the hydrogen bond H…O distance along the simulation. The shorter of the distances between the side chain NH hydrogen of N128 and any of the side chain O atoms of D96 is represented. Hydrogen bond breaking and reforming events are evidenced as peaks from the equilibrium distance baseline. (c) Statistics of hydrogen bond distances during a 4.5 ns simulation of wild type apoflavodoxin. Counts of distances sampled every ps are shown. The main peak represents the conformations that retain the hydrogen bond (see inset) while the flatter, wider peak represents conformations with a broken hydrogen bond. A 2.5 ± 0.1 Å cutoff has been used to calculate the free energy of hydrogen bond formation from the folded state.

**Figure 6**. Urea denaturation curves of wild type (solid circles) and D96A/N128A (open circles) apoflavodoxin double mutant (a) and of the D96A (solid circles) and N128A (open circles) single mutants (b). Data were recorded at 25.0 ± 0.1 °C in mops 50 mM pH 7.0 with 0.5 M NaCl, and fitted to a two-state equation (Santoro and Bolen, 1988).

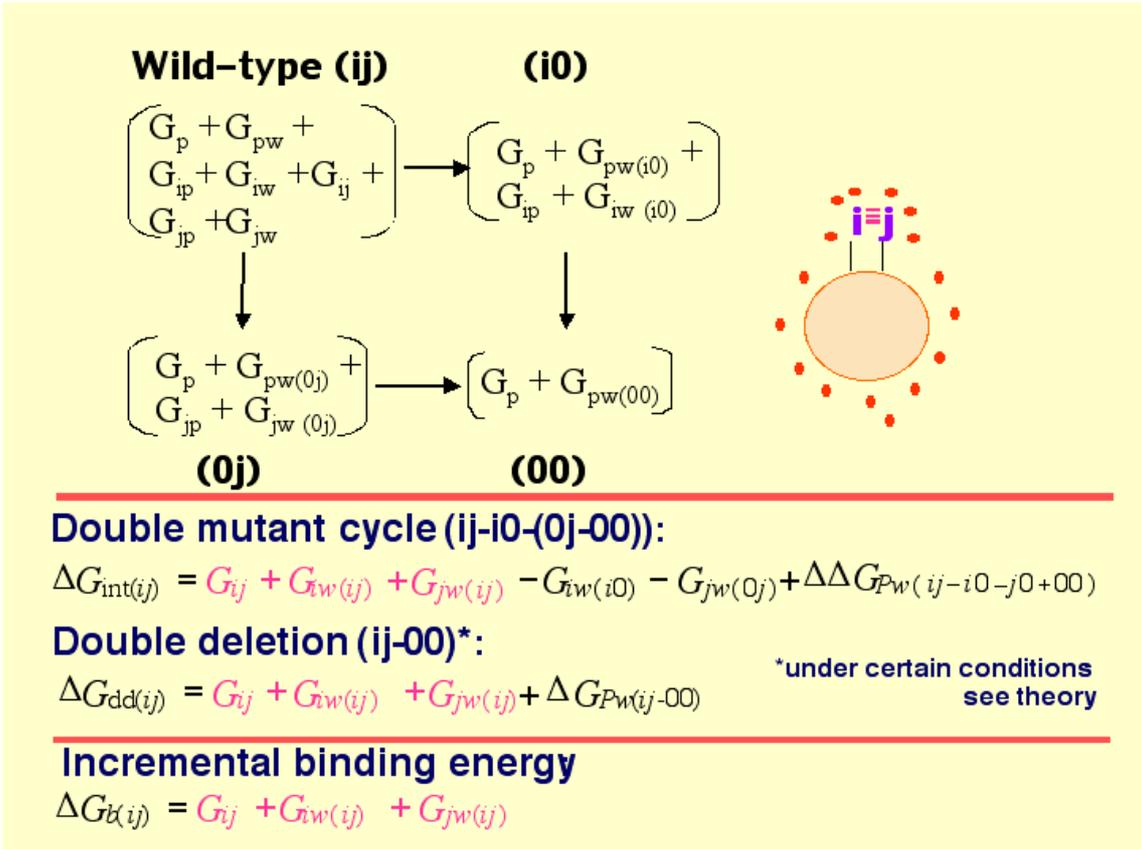

**Figure 1**

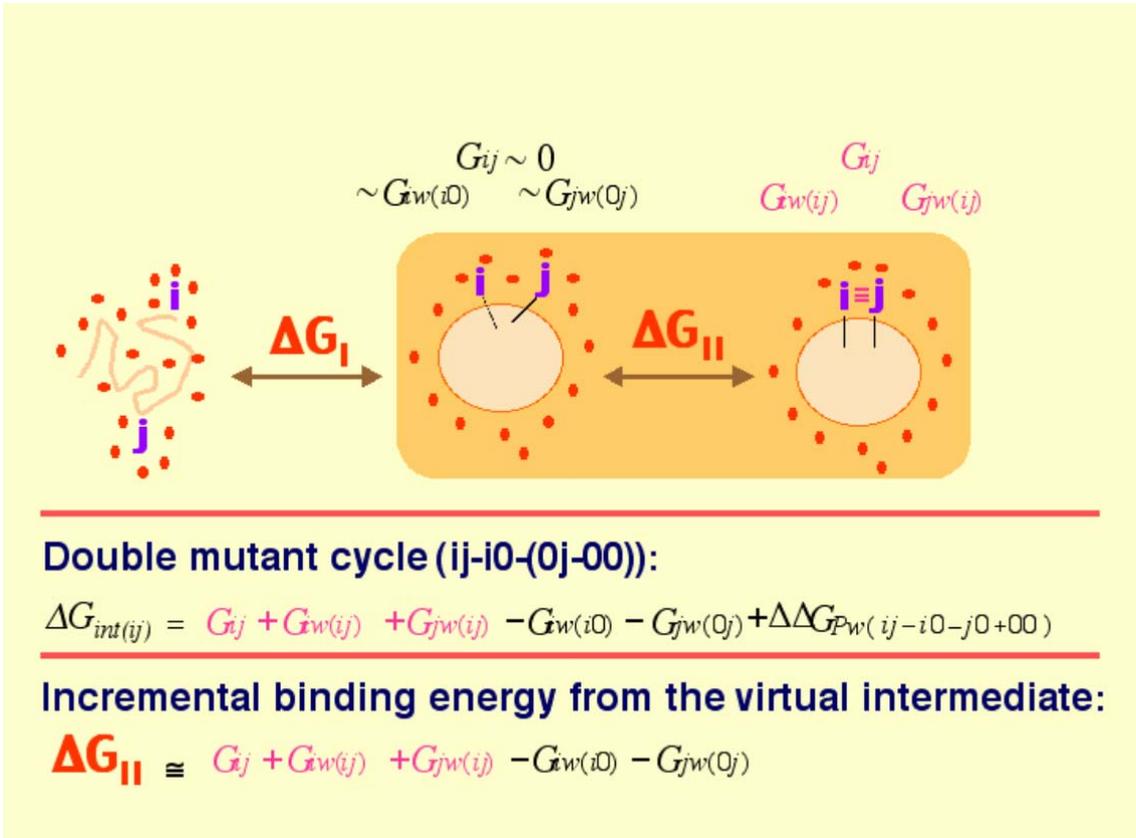

**Fig 2**

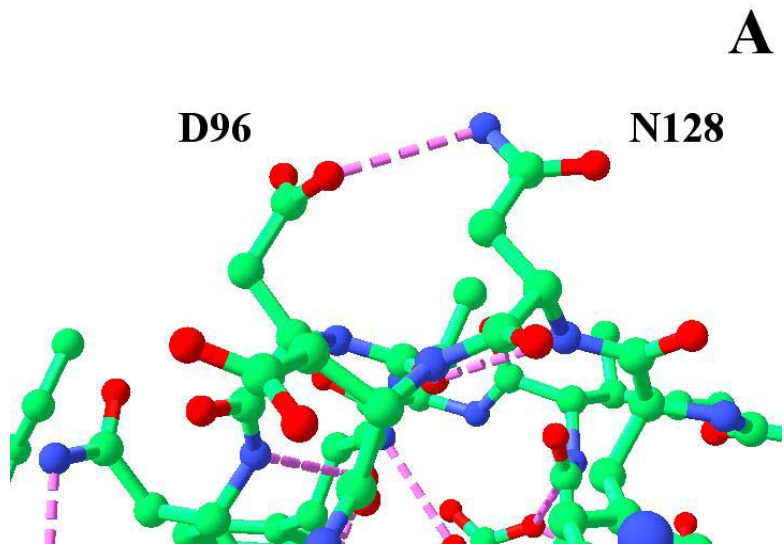

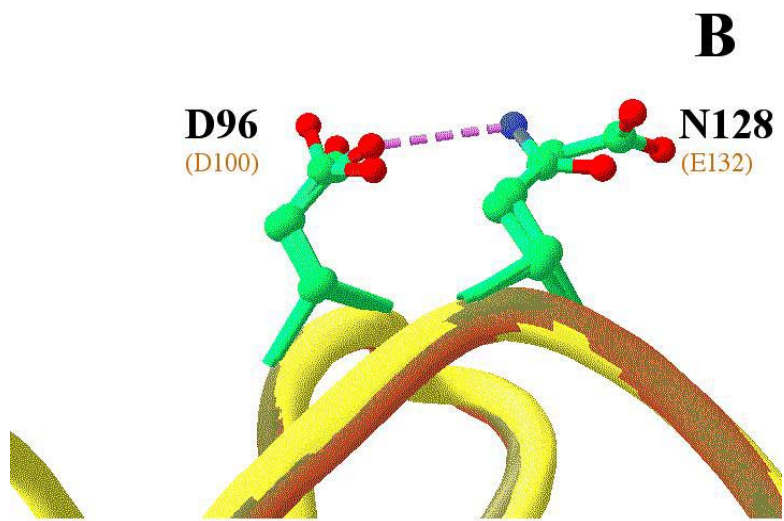

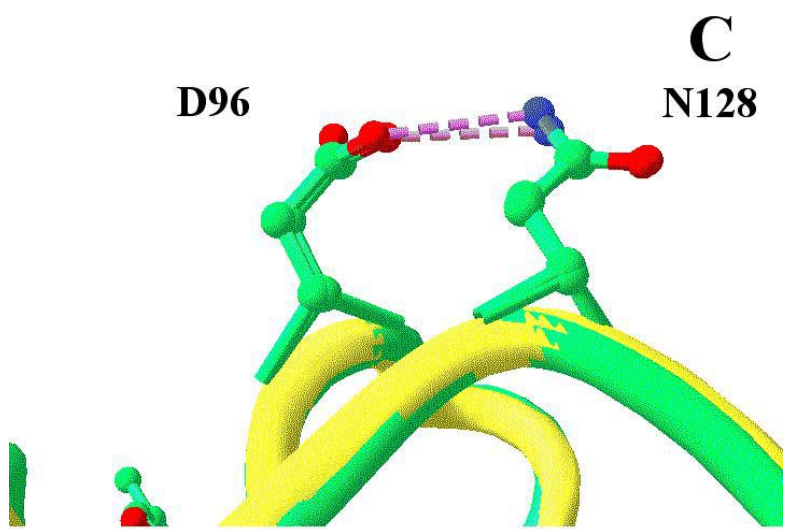

**Figure 3**

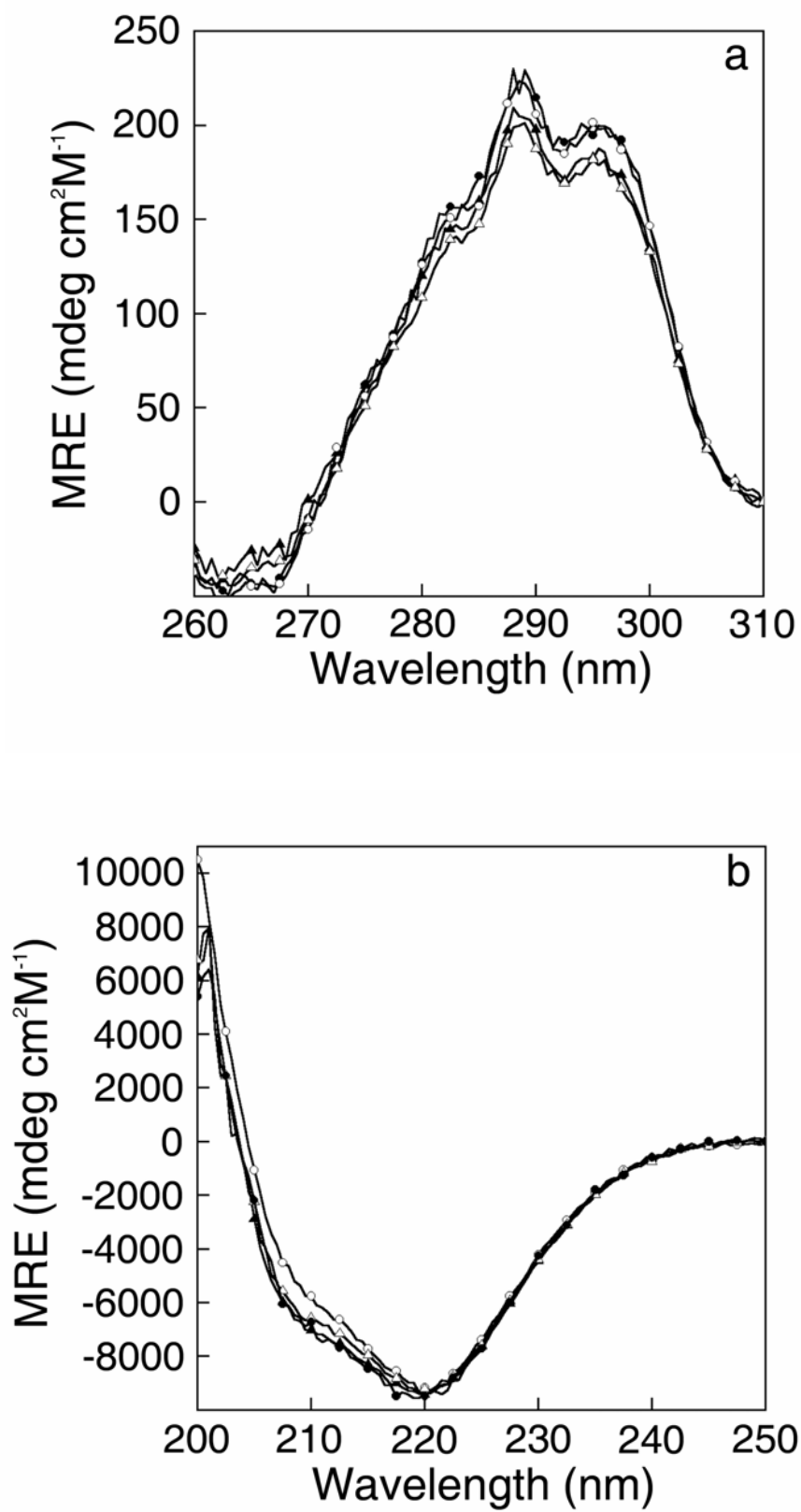

**Figure 4**

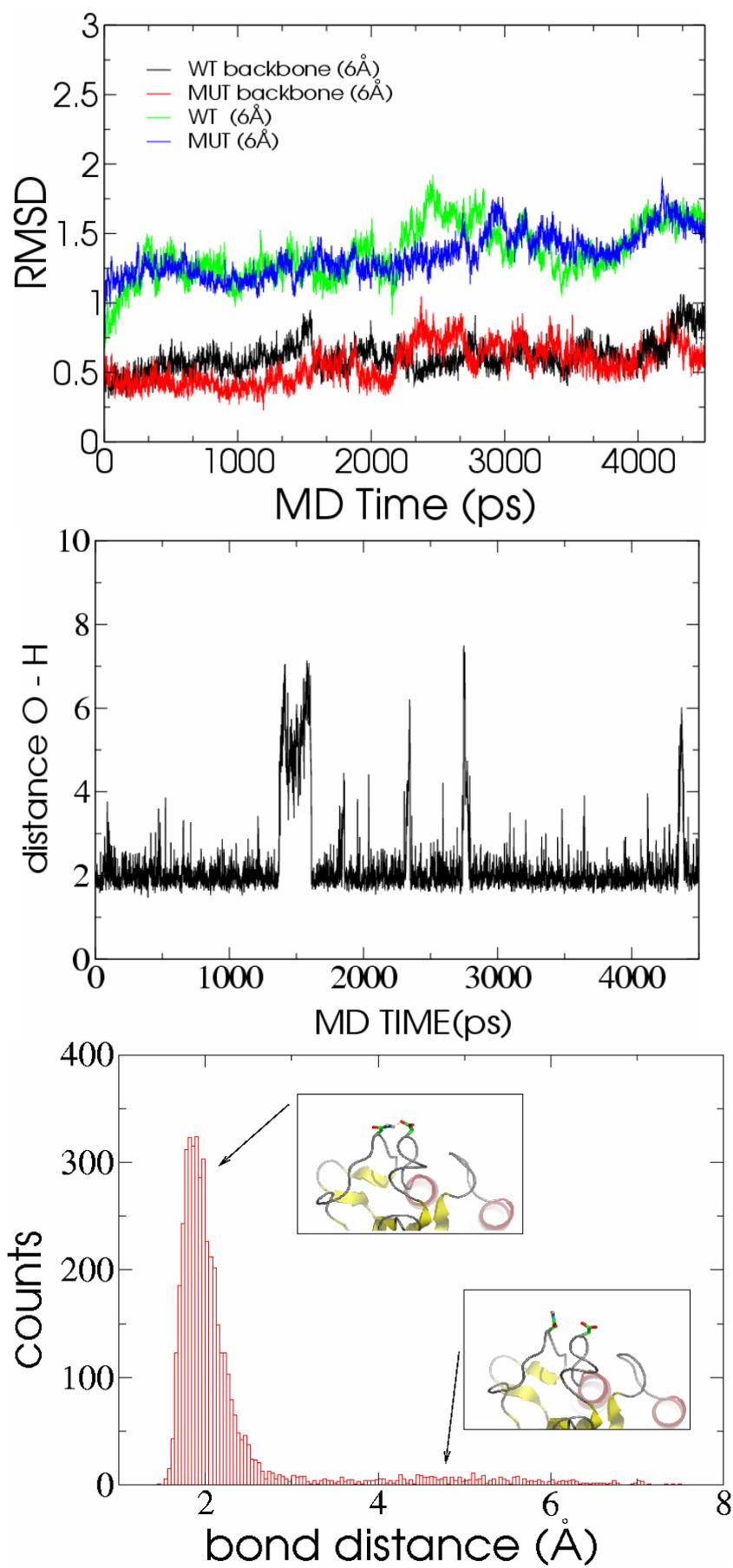

**Figure 5**

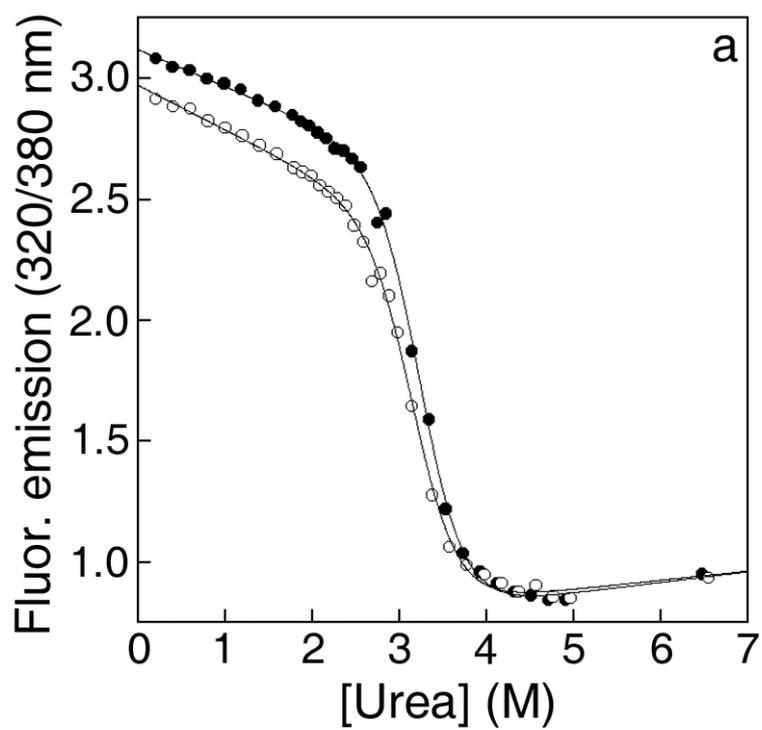

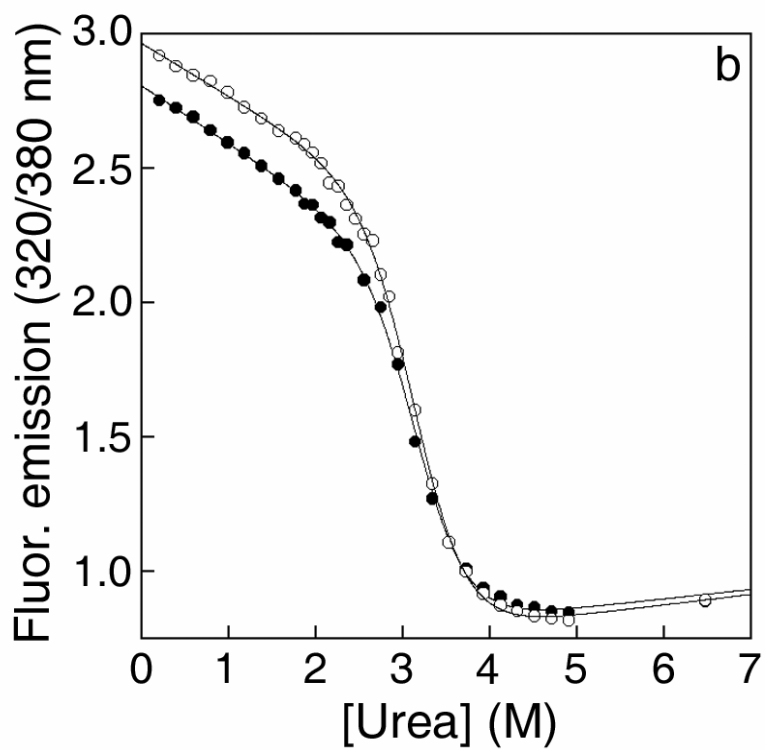

**Figure 6**